\documentclass[endfloats,12pt,aps,preprint,showpacs]{revtex4}
\usepackage{amsmath}
\usepackage{graphicx}

\begin{document}

\date{\today }
\title{Experimental evidence of percolation phase transition\\in surface plasmons generation} 
\author{Jean-Baptiste Masson}
\author{Guilhem Gallot}
\email[Corresponding author: ]{jeanbaptiste.masson@gmail.com}
\affiliation{Laboratoire d'Optique et Biosciences, \'{E}cole Polytechnique, CNRS UMR 7645 - INSERM U696, 91128 Palaiseau, France}

\begin{abstract}
Carrying digital information in traditional copper wires is becoming a major issue in electronic circuits. Optical connections such as fiber optics offers unprecedented transfer capacity, but the mismatch between the optical wavelength and the transistors size drastically reduces the coupling efficiency. By merging the abilities of photonics and electronics, surface plasmon photonics, or 'plasmonics' exhibits strong potential. Here, we propose an original approach to fully understand the nature of surface electrons in plasmonic systems, by experimentally demonstrating that surface plasmons can be modeled as a phase of surface waves. First and second order phase transitions, associated with percolation transitions, have been experimentally observed in the building process of surface plasmons in lattice of subwavelength apertures. Percolation theory provides a unified framework for surface plasmons description.
\end{abstract}


\maketitle 

Surface plasmons (SPs) plasmonics is of interest for a wide range of domains, from physics, chemistry, to biological sciences \cite{TK58,TK93}. SPs are associated with electromagnetic waves traveling along a metal/dielectric interface with collective motions of electrons. The tight localization of SPs at the surface of the plasmonic circuit provides the possibility of focusing and waveguiding light in subwavelength systems, which can closely interact with electronic devices. In the recent years, a renewed interest in SPs has arisen from the demonstration of a strong and unexpected enhancement of light transmission through arrays of sub-wavelength holes \cite{TK18}. Enhancement of several orders of magnitude has been reported with respect to standard aperture theory \cite{TK19,TK94,TK25}. The transmission can even exceed the surface ratio occupied by the holes, implying that light is focused by the structure of the arrays through the holes. These results are stimulating in numerous fields \cite{TK58,TK57,TK62,TK71,TK61}: near field microscopy, high density storage, detection of chemical and biological molecules, light emission, optical nonlinear devices.

Since this discovery, many studies have been performed to fully understand and control SPs, involving the influence of the shape, geometry lattice and polarization in one- and two-dimensional subwavelength structures \cite{TK16,TK50,TK67,TK15,TK96}. However, the fundamental question of the establishment of the SPs themselves has never been fully addressed yet. Here, we experimentally demonstrate that one- and two-dimensional SPs can be generated through a phase transition like process, separating phases of surface waves, i.e. specific organizations of the electromagnetic waves at the surface of the perforated metal plate. The analysis of the phase transitions features, and in particular the critical behavior found in 1D lattice SP establishment, leads to modeling by a percolation process. Percolation theory belongs to a universality class that unifies the description of many physical models. As a direct consequence of the established properties of percolation, this framework allows us a new insight of SPs, without the approach implying exact solving of Maxwell's equations.

The long-wavelength terahertz domain offers the advantage of the precise control of the geometry \cite{TK47}. Therefore, the filling parameter of the subwavelength structures is accurately calibrated. Furthermore, the realization of ultra-thin ($\lambda/50$) metal sheets removes possible wave-guiding Fabry-P\'{e}rot resonances \cite{TK53}. The structures are made of free-standing electroformed ultra-thin nickel plates, with precision of design better than 1~$\mu$m. Terahertz spectra are recorded using terahertz time-domain spectroscopy \cite{TA8}. Broadband linearly polarized subpicosecond single cycle pulses of terahertz radiation are generated and coherently detected by illuminating photoconductive antennas with two synchronized femtosecond laser pulses. The sample is positioned on a 10~mm circular aperture, in the linearly polarized, frequency independent, 4.8~mm-waist Gaussian THz beam ($1/e$ in amplitude). Numerical Fourier transform of the time-domain signals with and without the sample gives access to the transmission spectrum of the subwavelength structures. The dynamics of the surface plasmons is recorded over 240~ps, yielding a 3~GHz frequency precision after numerical Fourier transform, with 10$^4$ signal to noise ratio in 300~ms acquisition time.

We studied the generation of SPs by analyzing the terahertz transmission spectra of series of free-standing subwavelength 1D and 2D lattices (figures \ref{figure1}A-C) with increasing quantity of metal, while the periodicity remains constant. A filling parameter $p$ is defined as the ratio between the surface of the metal $S_m$ and the surface of the apertures $S_a$ in the lattice, by $p=S_m / S_a$, which enables to compare lattices of different shapes. SPs are built when increasing this filling parameter up to a precise level. Transmission spectra show typical resonant features, with minimum in 1D lattice (figure \ref{figure1}D), maximum in 2D lattice (figure \ref{figure1}E), and for both, a frequency shift $\Delta \nu=\nu - \nu_0$ of the first resonance is associated with the metal filling, where $\nu_0$ is the constant resonance frequency at low fillings. Indeed, in thin metal grids, resonant profiles due to diffraction over the periodic structure are observable near the transition and still allow defining a resonance frequency. Furthermore, we define the energy $U$ as the total electromagnetic energy blocked by the plate, and can be directly accessed from the transmission of the terahertz beam through the system $U=1-\int \left| A_t(t) \right| ^2 dt/\int \left| A_0(t) \right| ^2 dt$, where $A_t$ and $A_0$ are the measured amplitudes of the electric fields with and without the sample, respectively. We then define a geometric capacity $C$ linking the internal energy to the filling parameter, by $C=\partial U / \partial p$. We focused on the observation and evolution of these parameters, which provide fundamental insights on the nature of the generation of SPs.

Series of two-dimensional arrays of squared and round subwavelength holes (36 lattices each) have been analyzed (figure~\ref{figure1}B-C). Incident linear polarization is sent along direction $x$. At low metal fillings, transmission spectra (figure~\ref{figure1}E black) show two large resonance features with a factor 2 between the resonance frequencies and no frequency dependent shift. The incident electromagnetic wave is diffracted through the subwavelength apertures (Bethe's theory \cite{TK1}) and the transmitted waves interfere over the hole array. At high values of the filling parameter on the contrary, typical features of 2D lattice SPs can be observed (figure~\ref{figure1}E color) with sharp maxima and resonance frequencies scaling as $\sqrt 2$. A resonance frequency shift with $p$ is also observed, with precision better than 2~GHz. For tiny holes, first and second resonances agree well with theoretical resonance frequencies given by Bloch wave model at 0.5 and 0.71~THz, respectively \cite{TK19}, for a lattice period of $L=600~\mu$m. Increasing the filling parameter, experiments show clear discontinuity of the first resonance frequency and geometric capacity, at a specific transition filling parameter, both for squared (figure~\ref{figure2}A and C) and round holes (figure~\ref{figure2}B and D). This behavior is highly evocative of a first order phase transition. Transition filling parameter is the same for frequency shift and capacity, but different between squared and round holes, $p=1.03\pm0.01$ and $p=1.37\pm0.01$, respectively. It may also be related to a cut-off for the polariton propagation along the surface. Frequency shift is constant below the transition filling parameter and then monotonously increases after the discontinuity in the SP phase. It converges toward the value given for tiny holes, which is the same for squared and round holes \cite{TK18}. Frequency shift and capacity show different behavior whether the holes are squared or circular, with different transition filling parameter and discontinuity amplitude, confirming that SP is strongly influenced by the shape of the subwavelength holes. In addition, we have performed numerical simulations. We carried out a direct resolution of Maxwell's equations through a full three-dimensional (3D) \textit{ab initio} Finite Element Method (FEM) analysis of the electric field propagating trough the array \cite{TK96}. Results of the frequency shift for square and round apertures are shown in figure~\ref{figure2}, and demonstrate clear discontinuity at the same filing parameters found in the experiments.

The one-dimensional subwavelength lattices are made of very narrow slits (figure~\ref{figure1}A). Series of 1D subwavelength lattices of increasing filling has been realized with constant lattice period $L=$400~$\mu$m. The terahertz beam is sent perpendicularly to the lattice surface and linear polarization has an angle $\theta$ with respect to the transverse direction $y$. At low fillings, the subwavelength slits behave as standard polarizers. In particular, the spectra show a modulation with minimum resonances given by the diffraction over the metal lines (figure~\ref{figure1}D black), and the amplitude spectra scale proportionally to $\cos(\theta)$, with no resonance shift. On the contrary, at high fillings, the spectra show evidence of typical 1D SPs (figure~\ref{figure1}D red) with nearly zero minimum resonance of transmission (0.75~THz at $\theta=0^\circ$). Furthermore, this minimum shifts toward the low frequencies when incident polarization is rotated, due to the change of the projected apparent period. We measured the frequency shift for a set of 60 lattices of different filling parameters. For each angle $\theta$, the evolution of the frequency shift with the filling is characteristic of a second order phase transition (figure~\ref{figure3}A). The frequency shift remains constant in the low filling range, up to a critical value $p_c=1.583\pm0.005$, from which it continuously increases. This characteristic behavior is similar for all the incident polarizations. Evolution of the frequency shift versus the incident polarization angle shows sinusoidal variations (figure~\ref{figure3}B) as $\Delta \nu=\overline{\Delta \nu}\sin(\theta-\theta_0)$, where $\overline{\Delta \nu}$ is the intrinsic frequency shift and $\theta_0$ a possible small experimental angular misalignment. Precision on $\overline{\Delta \nu}$ is better than 1~GHz at maximum deviation. Defining a normalized filling parameter $\tau=(p-p_c)/p_c$, the evolution of the frequency shift in the SP domain near $p_c$ follows $\overline{\Delta \nu} \propto \left| \tau \right| ^\beta$, where $\beta$ is the critical coefficient associated with the frequency shift. Using $\overline{\Delta \nu}$, one obtains the critical coefficient $\beta=0.98\pm0.04$ (figure~\ref{figure3}C). Geometric capacity is also analyzed from internal energy. Capacity shows diverging features near $p_c$ (figure~\ref{figure4}). Geometric capacity evolves as $C \propto \left| \tau \right| ^\alpha$, where $\alpha$ is the critical coefficient associated with capacity. For both sides of $p_c$ (see inset in figure~\ref{figure4}), this critical coefficient has a value of $\alpha_+=-0.99\pm0.08$ and $\alpha_-=-0.97\pm0.08$, above and below $p_c$, respectively.

The analysis of the experimental data is now discussed. The main first point is that the data can not be fully understood by existing models of SPs. These models intrinsically consider the existence of only one possible state for the surface wave. Waveguide analysis or Bloch wave analysis provide an approximation of the resonance frequency \cite{LivreRaether} and give good values only in the limit case where the apertures are very tiny. Further refinements consist in decomposition on a basis, such as Rayleigh expansion, but their linear features fundamentally prohibit discontinuity. This discrepancy is illustrated from \cite{TK99} in figure~\ref{figure2} (dashed lines). On the contrary, we have observed here a first order transition behavior in 2D lattice, associated with the simultaneous discontinuity of both frequency shift and geometric capacity at the same filling value. Moreover, in 1D lattice, power law evolution of frequency shift and capacity is intimately linked to second order phase transition, and the critical coefficients are clearly in favor of a percolation process. Another strong hint is the polarization response below and above $p_c$. There is a spontaneous symmetry breaking phenomenon at $p_c$, since original isotropic response below $p_c$ is broken. Indeed, at low fillings, the shape of the spectra remains unchanged with the rotation of the incident polarization. On the contrary, the frequency resonance shifts with the polarization angle above $p_c$ (see figure~\ref{figure3}A), corresponding to a disappearing of the symmetry. Therefore, the generation of SPs can be described in a phase transition framework. The first phase ($p<p_c$) consists of non resonant surface waves diffracted by the lattice in a way similar to Bethe description \cite{TK1}. The second phase ($p>p_c$) is SPs characterized by a resonant interaction with the lattice.

Complex systems can be described by statistical physics, in particular systems involving geometrical consideration, such as percolation in conductivity, polymerization or forest fire propagation. In the same way, in ferromagnetic/paramagnetic transitions, only the measurement of the linear magnetic susceptibility gives access to the second order phase transition. Phase transitions imply local non linear interactions. Even though the measured quantities are linear with respect to the electromagnetic wave going through the plate, the variation of the size of the aperture changes the relative contribution of far- and near-field interactions, which behave as $1/r$ and $1/r^3$. Thus, non linear features with respect to the geometry are the key of the understanding of our system. Furthermore, the screen transmission is not linear with the number of holes, since the transmitted signal is not the sum of the signal of a single hole.

From macroscopic measurements, we demonstrated the phase transition origin of SP generation. Universality class theory shows that the critical exponents for the phase transitions for broad classes of physical systems are the same. Therefore, a model describing a specific system can be extended to another system within its universality class. Then, experimental critical values of $\alpha$ and $\beta$, associated with universality of percolation theory provide us a microscopic description of SP generation. The first order phase transition is also described by percolation and the difference between 1D and 2D lattices originate from the change of degree of freedom in the system. Percolation transition describes microscopic ``sites'' linked together by ``bonds''. Consequences of percolation theory are that in SPs the sites are the (randomly moving) surface electrons, and the bonds are the coupling between the electrons and photons from the electromagnetic wave. Surface electrons are locally correlated by the interaction with the electromagnetic wave, defining a cluster of electrons as an ensemble of correlated surface electrons. A percolation transition then occurs when the first cluster reaches the size of the screen. At low filling factor, the size of the clusters is small, electrons only interact locally, and no SPs are present (cf figure~\ref{figure1}D-E). When $p>p_c$, large groups of surface electrons interact together, forming clusters of macroscopic size and establishing a SP. Furthermore, the experimental critical parameters $\alpha \approx -1$ and $\beta \approx 1$ can be related to the percolation class of Cayley tree \cite{LivreSchwabl}, which provides the critical evolution of the cluster size, as $\chi \propto \left| p-p_c \right|^{-\gamma}$, with $\gamma=1$.

It has been demonstrated that SPs are a phase of surface electrons coupled to photons, and that SP building implies percolation phase transitions with respect to a filling parameter. In 2D lattice, a first order phase transition exhibits discontinuity of order parameter and geometric capacity, associated with a hole shape dependent transition filling parameter. In 1D lattice, second order phase transition was analyzed and two critical coefficients measured, associated with Bethe lattice percolation model. Under critical conditions, correlation length diverges, with the consequence that information can be coherently transmitted through the plasmonic device on very large distance. Furthermore, 1D SP lattice appears to be promising easy-to-handle experimental tools for studying second order phase transitions.

\newpage

\begin{figure}[f]
\begin{center}
\end{center}
\caption{\textbf{Geometry} of the 1D \textbf{(A)} and 2D \textbf{(B-C)} lattices. \textbf{Terahertz transmission spectra.} \textbf{(D)} 1D lattice spectra, at low filling ($p=1.5$, black line) and high filling (T=1.67, $\theta=0^\circ$ solid red line, $\theta=80^\circ$ dashed red line). \textbf{(E)} 2D lattice spectra of squared holes, at low filling ($p=0.78$, black line) and high filling ($p=3.3$ green line, $p=7.2$ blue line, $p=9$ red line). Arrows show theoretical Bloch wave theory frequencies.}
\label{figure1}
\end{figure}

\begin{figure}[f]
\begin{center}
\end{center}
\caption{\textbf{First order phase transition in 2D lattices.} Evolution of the frequency shift for squared holes \textbf{(A)} and round holes \textbf{(B)}. Solid lines are 3D FEM numerical simulations of the evolution of the frequency shift. Evolution of geometric capacity for squared holes \textbf{(C)} and round holes \textbf{(D)}. Transition filling parameters for squared and round holes are $p=1.03\pm0.01$ and $p=1.37\pm0.01$, respectively. Dashed lines are linear calculations from \cite{TK99}.}
\label{figure2}
\end{figure}

\begin{figure}[f]
\begin{center}
\end{center}
\caption{\textbf{Second order phase transition in 1D lattice: resonance frequency shift.} \textbf{(A)} Evolution of the frequency shift $\Delta \nu$ with filling parameter and corresponding slit width, for incident polarization angle $\theta=$80$^\circ$ (black), 45$^\circ$ (red) and 20$^\circ$ (blue). Critical filling parameter is $p_c=1.583\pm0.005$.  \textbf{(B)} Evolution of the frequency shift with incident polarization at $p=1.67$ for experimental data (solid circles), and sinusoidal fit (solid line). \textbf{(C)} Evolution of the frequency shift versus normalized filling parameter in logarithm scale showing the critical behaviour for data (circles) and critical evolution fit with $\beta=0.98\pm0.04$ (solid line).}
\label{figure3}
\end{figure}

\begin{figure}[f]
\begin{center}
\end{center}
\caption{\textbf{Second order phase transition in 1D lattice: geometric capacity.} Evolution of capacity versus filling parameter for experimental data (circles) and critical divergence fit (solid lines). Inset shows critical evolution at both sides of the critical filling parameter, with $\alpha_+ = -0.99 \pm0.05$ and $\alpha_-=-0.97\pm0.05$, above and below $p_c$, respectively.}
\label{figure4}
\end{figure}

\printfigures

\newpage
\begin{center}
\includegraphics[width=450pt]{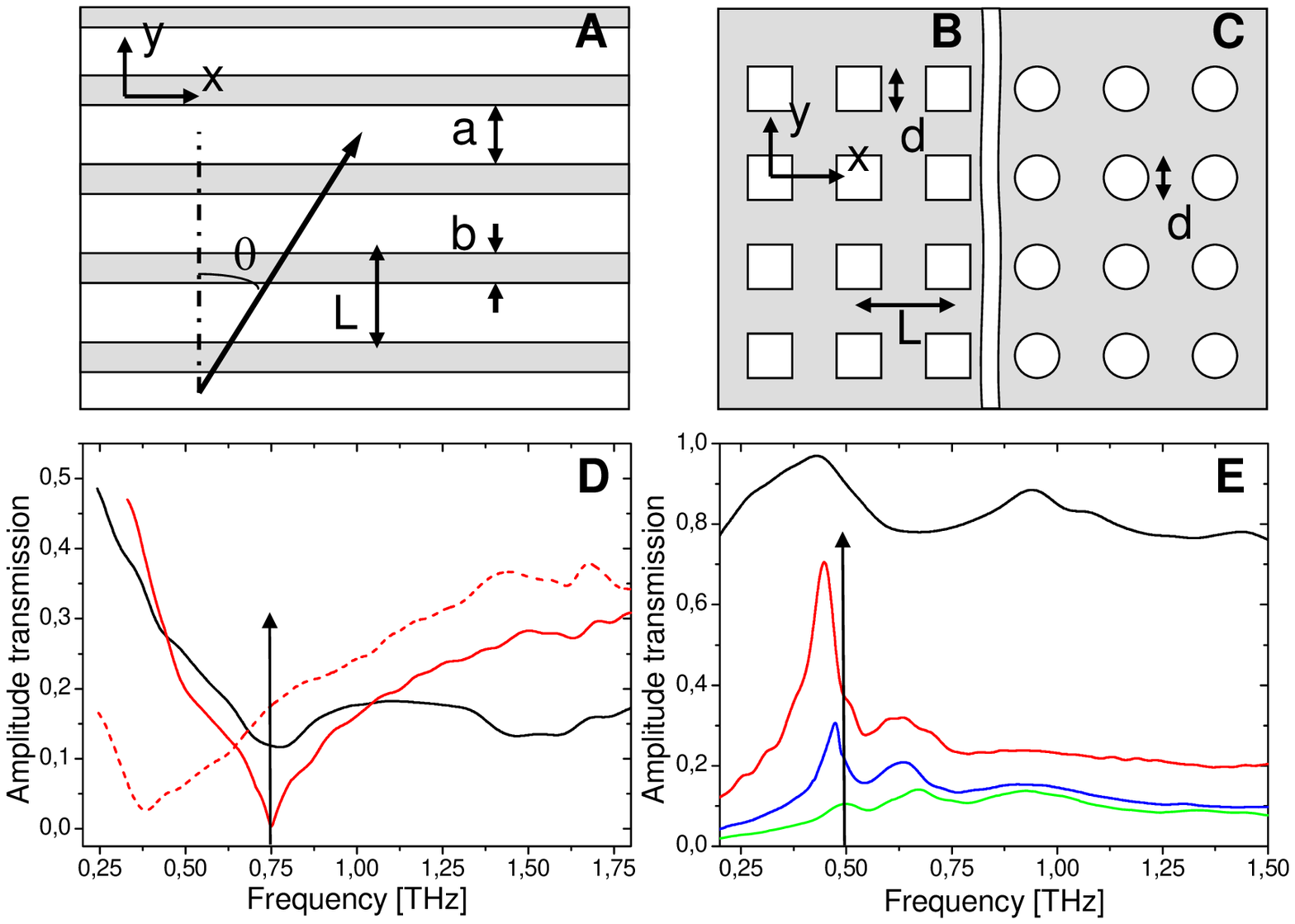}

\vspace{4cm} Masson et al. Figure \ref{figure1}
\end{center}

\newpage
\begin{center}
\includegraphics[width=15cm]{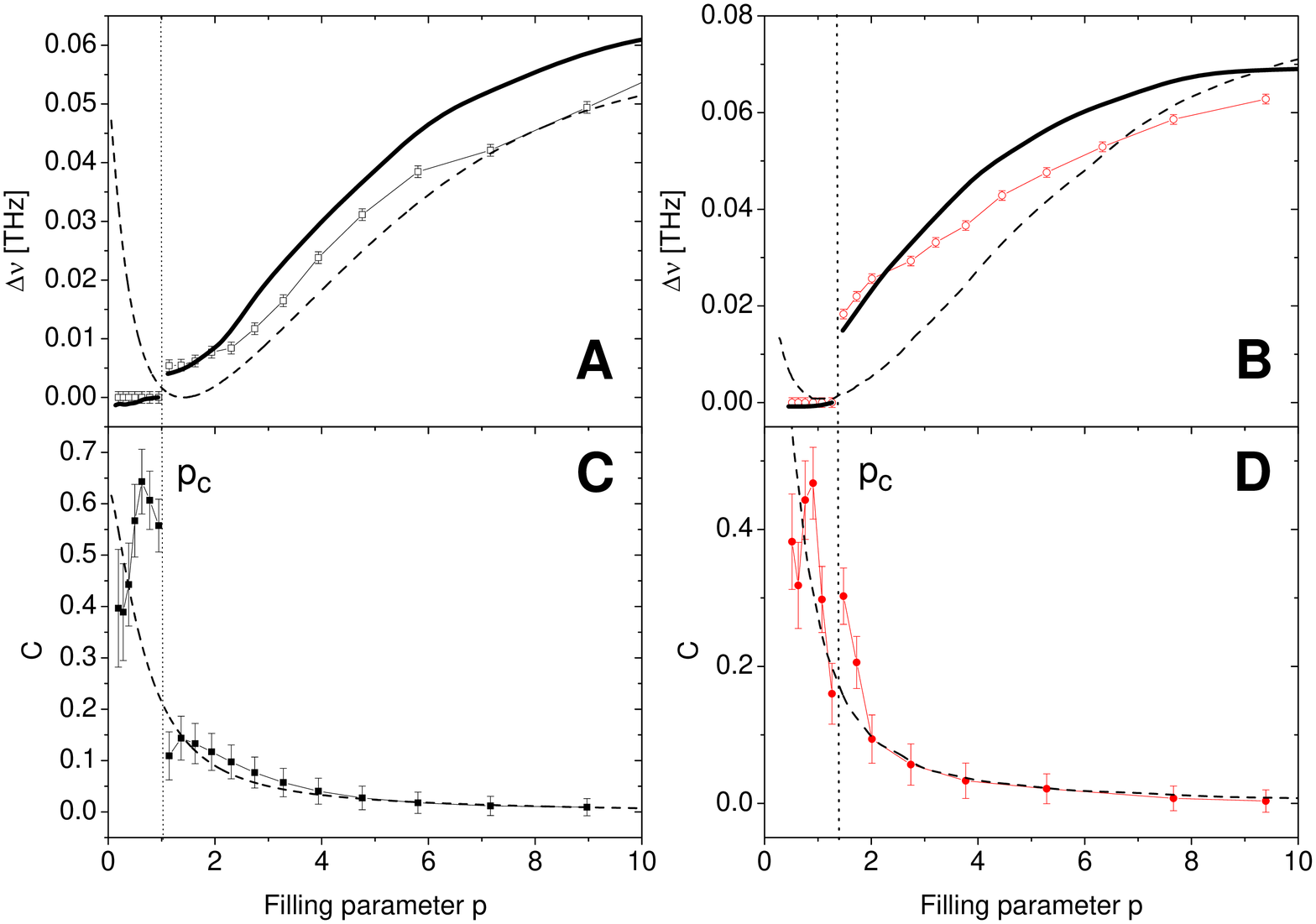}

\vspace{2cm} Masson et al. Figure \ref{figure2}
\end{center}

\newpage
\begin{center}
\includegraphics[width=260pt]{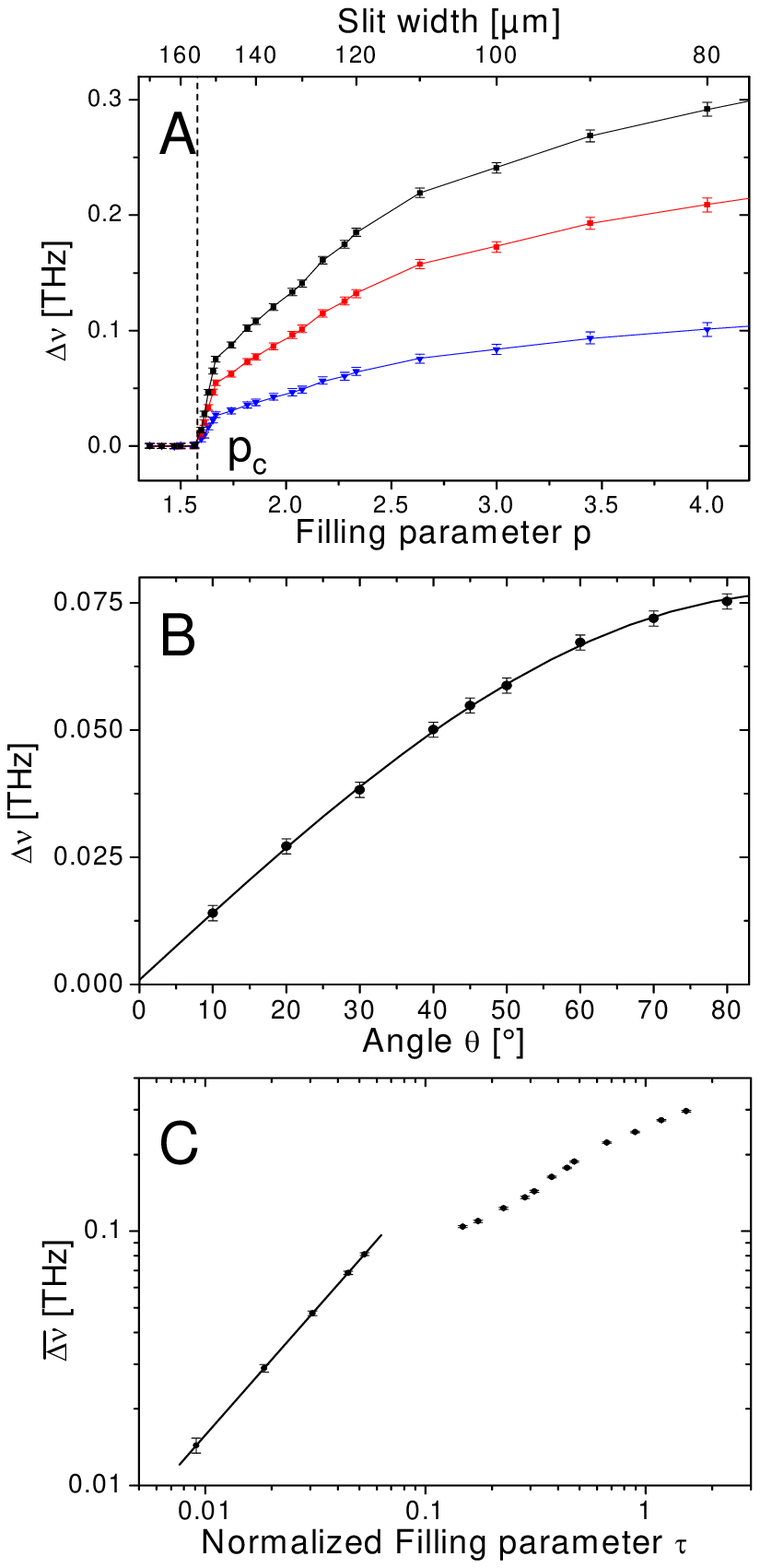}

\vspace{0cm} Masson et al. Figure \ref{figure3}
\end{center}

\newpage
\begin{center}
\includegraphics[width=400pt]{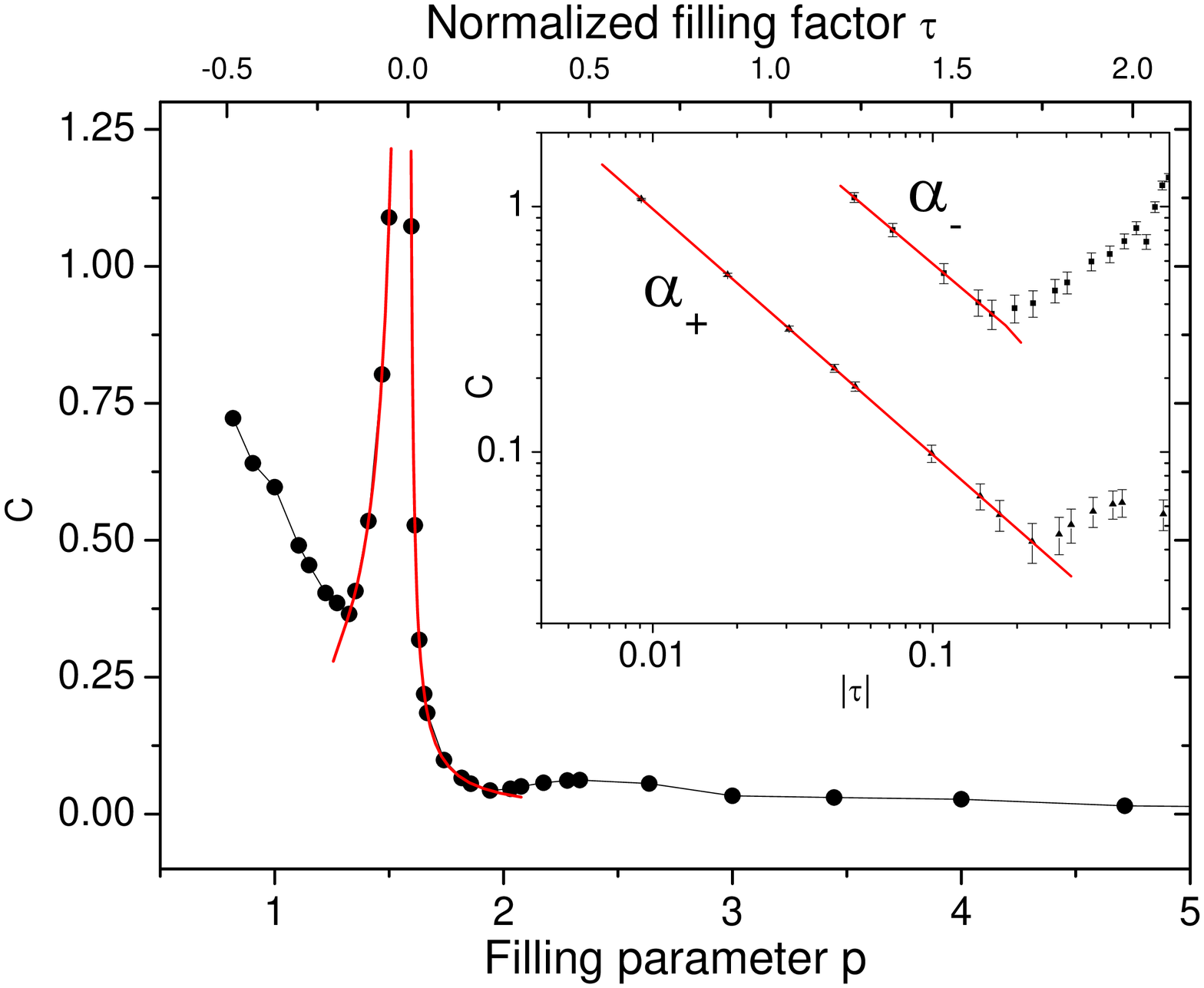}

\vspace{2cm} Masson et al. Figure \ref{figure4}
\end{center}


\newpage

\begin{thebibliography}{22}
\expandafter\ifx\csname natexlab\endcsname\relax\def\natexlab#1{#1}\fi
\expandafter\ifx\csname bibnamefont\endcsname\relax
  \def\bibnamefont#1{#1}\fi
\expandafter\ifx\csname bibfnamefont\endcsname\relax
  \def\bibfnamefont#1{#1}\fi
\expandafter\ifx\csname citenamefont\endcsname\relax
  \def\citenamefont#1{#1}\fi
\expandafter\ifx\csname url\endcsname\relax
  \def\url#1{\texttt{#1}}\fi
\expandafter\ifx\csname urlprefix\endcsname\relax\def\urlprefix{URL }\fi
\providecommand{\bibinfo}[2]{#2}
\providecommand{\eprint}[2][]{\url{#2}}

\bibitem[{\citenamefont{Barnes et~al.}(2003)\citenamefont{Barnes, Dereux, and
  Ebbesen}}]{TK58}
\bibinfo{author}{\bibfnamefont{W.~L.} \bibnamefont{Barnes}},
  \bibinfo{author}{\bibfnamefont{A.}~\bibnamefont{Dereux}}, \bibnamefont{and}
  \bibinfo{author}{\bibfnamefont{T.~W.} \bibnamefont{Ebbesen}},
  \bibinfo{journal}{Nature} \textbf{\bibinfo{volume}{424}},
  \bibinfo{pages}{824} (\bibinfo{year}{2003}).

\bibitem[{\citenamefont{Ozbay}(2006)}]{TK93}
\bibinfo{author}{\bibfnamefont{E.}~\bibnamefont{Ozbay}},
  \bibinfo{journal}{Science} \textbf{\bibinfo{volume}{311}},
  \bibinfo{pages}{189} (\bibinfo{year}{2006}).

\bibitem[{\citenamefont{Ebbesen et~al.}(1998)\citenamefont{Ebbesen, Lezec,
  Ghaemi, Thio, and Wolff}}]{TK18}
\bibinfo{author}{\bibfnamefont{T.~W.} \bibnamefont{Ebbesen}},
  \bibinfo{author}{\bibfnamefont{H.~J.} \bibnamefont{Lezec}},
  \bibinfo{author}{\bibfnamefont{H.~F.} \bibnamefont{Ghaemi}},
  \bibinfo{author}{\bibfnamefont{T.}~\bibnamefont{Thio}}, \bibnamefont{and}
  \bibinfo{author}{\bibfnamefont{P.~A.} \bibnamefont{Wolff}},
  \bibinfo{journal}{Nature} \textbf{\bibinfo{volume}{391}},
  \bibinfo{pages}{667} (\bibinfo{year}{1998}).

\bibitem[{\citenamefont{Martin-Moreno et~al.}(2001)\citenamefont{Martin-Moreno,
  Garcia-Vidal, Lezec, Pellerin, Thio, Pendry, and Ebbesen}}]{TK19}
\bibinfo{author}{\bibfnamefont{L.}~\bibnamefont{Martin-Moreno}},
  \bibinfo{author}{\bibfnamefont{F.~J.} \bibnamefont{Garcia-Vidal}},
  \bibinfo{author}{\bibfnamefont{H.~J.} \bibnamefont{Lezec}},
  \bibinfo{author}{\bibfnamefont{K.~M.} \bibnamefont{Pellerin}},
  \bibinfo{author}{\bibfnamefont{T.}~\bibnamefont{Thio}},
  \bibinfo{author}{\bibfnamefont{J.~B.} \bibnamefont{Pendry}},
  \bibnamefont{and} \bibinfo{author}{\bibfnamefont{T.~W.}
  \bibnamefont{Ebbesen}}, \bibinfo{journal}{Phys. Rev. Lett.}
  \textbf{\bibinfo{volume}{86}}, \bibinfo{pages}{1114} (\bibinfo{year}{2001}).

\bibitem[{\citenamefont{Porto et~al.}(1999)\citenamefont{Porto, Garcia-Vidal,
  and Pendry}}]{TK94}
\bibinfo{author}{\bibfnamefont{J.~A.} \bibnamefont{Porto}},
  \bibinfo{author}{\bibfnamefont{F.~J.} \bibnamefont{Garcia-Vidal}},
  \bibnamefont{and} \bibinfo{author}{\bibfnamefont{J.~B.}
  \bibnamefont{Pendry}}, \bibinfo{journal}{Phys. Rev. Lett.}
  \textbf{\bibinfo{volume}{83}}, \bibinfo{pages}{2845} (\bibinfo{year}{1999}).

\bibitem[{\citenamefont{Hibbins et~al.}(2005)\citenamefont{Hibbins, Evans, and
  Sambles}}]{TK25}
\bibinfo{author}{\bibfnamefont{A.~P.} \bibnamefont{Hibbins}},
  \bibinfo{author}{\bibfnamefont{B.~R.} \bibnamefont{Evans}}, \bibnamefont{and}
  \bibinfo{author}{\bibfnamefont{J.~R.} \bibnamefont{Sambles}},
  \bibinfo{journal}{Science} \textbf{\bibinfo{volume}{308}},
  \bibinfo{pages}{670} (\bibinfo{year}{2005}).

\bibitem[{\citenamefont{Liu and Blair}(2003)}]{TK57}
\bibinfo{author}{\bibfnamefont{Y.}~\bibnamefont{Liu}} \bibnamefont{and}
  \bibinfo{author}{\bibfnamefont{S.}~\bibnamefont{Blair}},
  \bibinfo{journal}{Opt. Lett.} \textbf{\bibinfo{volume}{28}},
  \bibinfo{pages}{507} (\bibinfo{year}{2003}).

\bibitem[{\citenamefont{Sambles}(1998)}]{TK62}
\bibinfo{author}{\bibfnamefont{R.}~\bibnamefont{Sambles}},
  \bibinfo{journal}{Nature} \textbf{\bibinfo{volume}{391}},
  \bibinfo{pages}{641} (\bibinfo{year}{1998}).

\bibitem[{\citenamefont{Vuckovic et~al.}(2000)\citenamefont{Vuckovic, Loncar,
  and Scherer}}]{TK71}
\bibinfo{author}{\bibfnamefont{J.}~\bibnamefont{Vuckovic}},
  \bibinfo{author}{\bibfnamefont{M.}~\bibnamefont{Loncar}}, \bibnamefont{and}
  \bibinfo{author}{\bibfnamefont{A.}~\bibnamefont{Scherer}},
  \bibinfo{journal}{IEEE J. Quantum Electron.} \textbf{\bibinfo{volume}{36}},
  \bibinfo{pages}{1131} (\bibinfo{year}{2000}).

\bibitem[{\citenamefont{Altewischer et~al.}(2002)\citenamefont{Altewischer,
  Exter, and Woerdman}}]{TK61}
\bibinfo{author}{\bibfnamefont{E.}~\bibnamefont{Altewischer}},
  \bibinfo{author}{\bibfnamefont{M.~P.~v.} \bibnamefont{Exter}},
  \bibnamefont{and} \bibinfo{author}{\bibfnamefont{J.~P.}
  \bibnamefont{Woerdman}}, \bibinfo{journal}{Nature}
  \textbf{\bibinfo{volume}{418}}, \bibinfo{pages}{304} (\bibinfo{year}{2002}).

\bibitem[{\citenamefont{Pendry et~al.}(2004)\citenamefont{Pendry,
  Martin-Moreno, and Garcia-Vidal}}]{TK16}
\bibinfo{author}{\bibfnamefont{J.~B.} \bibnamefont{Pendry}},
  \bibinfo{author}{\bibfnamefont{L.}~\bibnamefont{Martin-Moreno}},
  \bibnamefont{and} \bibinfo{author}{\bibfnamefont{F.~J.}
  \bibnamefont{Garcia-Vidal}}, \bibinfo{journal}{Science}
  \textbf{\bibinfo{volume}{305}}, \bibinfo{pages}{847} (\bibinfo{year}{2004}).

\bibitem[{\citenamefont{Gordon et~al.}(2004)\citenamefont{Gordon, Brolo,
  McKinnon, Rajora, Leathem, and Kavanagh}}]{TK50}
\bibinfo{author}{\bibfnamefont{R.}~\bibnamefont{Gordon}},
  \bibinfo{author}{\bibfnamefont{A.~G.} \bibnamefont{Brolo}},
  \bibinfo{author}{\bibfnamefont{A.}~\bibnamefont{McKinnon}},
  \bibinfo{author}{\bibfnamefont{A.}~\bibnamefont{Rajora}},
  \bibinfo{author}{\bibfnamefont{B.}~\bibnamefont{Leathem}}, \bibnamefont{and}
  \bibinfo{author}{\bibfnamefont{K.~L.} \bibnamefont{Kavanagh}},
  \bibinfo{journal}{Phys. Rev. Lett.} \textbf{\bibinfo{volume}{92}},
  \bibinfo{pages}{037401} (\bibinfo{year}{2004}).

\bibitem[{\citenamefont{Kim et~al.}(2003)\citenamefont{Kim, Hohng, Malyarchuk,
  Yoon, Ahn, Yee, Park, Kim, Park, and Lienau}}]{TK67}
\bibinfo{author}{\bibfnamefont{D.~S.} \bibnamefont{Kim}},
  \bibinfo{author}{\bibfnamefont{S.~C.} \bibnamefont{Hohng}},
  \bibinfo{author}{\bibfnamefont{V.}~\bibnamefont{Malyarchuk}},
  \bibinfo{author}{\bibfnamefont{Y.~C.} \bibnamefont{Yoon}},
  \bibinfo{author}{\bibfnamefont{Y.~H.} \bibnamefont{Ahn}},
  \bibinfo{author}{\bibfnamefont{K.~J.} \bibnamefont{Yee}},
  \bibinfo{author}{\bibfnamefont{J.~W.} \bibnamefont{Park}},
  \bibinfo{author}{\bibfnamefont{J.}~\bibnamefont{Kim}},
  \bibinfo{author}{\bibfnamefont{Q.~H.} \bibnamefont{Park}}, \bibnamefont{and}
  \bibinfo{author}{\bibfnamefont{C.}~\bibnamefont{Lienau}},
  \bibinfo{journal}{Phys. Rev. Lett.} \textbf{\bibinfo{volume}{91}},
  \bibinfo{pages}{143901} (\bibinfo{year}{2003}).

\bibitem[{\citenamefont{Koerkamp et~al.}(2004)\citenamefont{Koerkamp, Enoch,
  Segerink, Hulst, and Kuipers}}]{TK15}
\bibinfo{author}{\bibfnamefont{K.~J.~K.} \bibnamefont{Koerkamp}},
  \bibinfo{author}{\bibfnamefont{S.}~\bibnamefont{Enoch}},
  \bibinfo{author}{\bibfnamefont{F.~B.} \bibnamefont{Segerink}},
  \bibinfo{author}{\bibfnamefont{N.~F.~v.} \bibnamefont{Hulst}},
  \bibnamefont{and} \bibinfo{author}{\bibfnamefont{L.}~\bibnamefont{Kuipers}},
  \bibinfo{journal}{Phys. Rev. Lett.} \textbf{\bibinfo{volume}{92}},
  \bibinfo{pages}{183901} (\bibinfo{year}{2004}).

\bibitem[{\citenamefont{Masson and Gallot}(2006)}]{TK96}
\bibinfo{author}{\bibfnamefont{J.-B.} \bibnamefont{Masson}} \bibnamefont{and}
  \bibinfo{author}{\bibfnamefont{G.}~\bibnamefont{Gallot}},
  \bibinfo{journal}{Phys. Rev. B} \textbf{\bibinfo{volume}{73}},
  \bibinfo{pages}{121401(R)} (\bibinfo{year}{2006}).

\bibitem[{\citenamefont{Qu and Grischkowsky}(2004)}]{TK47}
\bibinfo{author}{\bibfnamefont{D.}~\bibnamefont{Qu}} \bibnamefont{and}
  \bibinfo{author}{\bibfnamefont{D.}~\bibnamefont{Grischkowsky}},
  \bibinfo{journal}{Phys. Rev. Lett.} \textbf{\bibinfo{volume}{93}},
  \bibinfo{pages}{196804} (\bibinfo{year}{2004}).

\bibitem[{\citenamefont{Cao and Lalanne}(2002)}]{TK53}
\bibinfo{author}{\bibfnamefont{Q.}~\bibnamefont{Cao}} \bibnamefont{and}
  \bibinfo{author}{\bibfnamefont{P.}~\bibnamefont{Lalanne}},
  \bibinfo{journal}{Phys. Rev. Lett.} \textbf{\bibinfo{volume}{88}},
  \bibinfo{pages}{057403} (\bibinfo{year}{2002}).

\bibitem[{\citenamefont{Grischkowsky et~al.}(1990)\citenamefont{Grischkowsky,
  Keiding, van Exter, and Fattinger}}]{TA8}
\bibinfo{author}{\bibfnamefont{D.}~\bibnamefont{Grischkowsky}},
  \bibinfo{author}{\bibfnamefont{S.~R.} \bibnamefont{Keiding}},
  \bibinfo{author}{\bibfnamefont{M.}~\bibnamefont{van Exter}},
  \bibnamefont{and}
  \bibinfo{author}{\bibfnamefont{C.}~\bibnamefont{Fattinger}},
  \bibinfo{journal}{J. Opt. Soc. Am. B} \textbf{\bibinfo{volume}{7}},
  \bibinfo{pages}{2006} (\bibinfo{year}{1990}).

\bibitem[{\citenamefont{Bethe}(1944)}]{TK1}
\bibinfo{author}{\bibfnamefont{H.~A.} \bibnamefont{Bethe}},
  \bibinfo{journal}{Phys. Rev.} \textbf{\bibinfo{volume}{66}},
  \bibinfo{pages}{163} (\bibinfo{year}{1944}).

\bibitem[{\citenamefont{Raether}(1988)}]{LivreRaether}
\bibinfo{author}{\bibfnamefont{H.}~\bibnamefont{Raether}},
  \emph{\bibinfo{title}{Surface Plasmons}}
  (\bibinfo{publisher}{Spinger-Verlag}, \bibinfo{address}{Berlin},
  \bibinfo{year}{1988}).

\bibitem[{\citenamefont{Chen}(1973)}]{TK99}
\bibinfo{author}{\bibfnamefont{C.-C.} \bibnamefont{Chen}},
  \bibinfo{journal}{IEEE trans. microwave theo. tech.}
  \textbf{\bibinfo{volume}{21}}, \bibinfo{pages}{1} (\bibinfo{year}{1973}).

\bibitem[{\citenamefont{Schwabl}(2002)}]{LivreSchwabl}
\bibinfo{author}{\bibfnamefont{F.}~\bibnamefont{Schwabl}},
  \emph{\bibinfo{title}{Statistical mechanics}} (\bibinfo{publisher}{Spinger},
  \bibinfo{address}{Berlin}, \bibinfo{year}{2002}).

\end{thebibliography}
\end{document}